**RESEARCH HIGHLIGHTS**

- The database used is the Airline On-Time Performance Data from the Bureau of Transport Statistics of the US.

- The goal is to provide a global picture of the statistical patterns of the reactionary delays with the aim of developing models able to realistically account for the delay propagation.

- We study topological properties of the US air transportation network in 2010.

- The delay distributions, for both arrival and departure, show long decays which is a signature of the complex nature of the phenomena taking place in the system.

- The aircraft rotation shows a highly heterogeneous profile that affects the development and propagation of delays.

- For flights with delay longer than 12 hours, the destination airport plays an important role.

# CHARACTERIZATION OF DELAY PROPAGATION IN THE US AIR TRANSPORTATION NETWORK

Pablo Fleurquin[1,2,*], José J. Ramasco[1], Víctor M. Eguíluz[1]

[1]Instituto de Física Interdisciplinaria y Sistemas Complejos IFISC (CSIC-UIB), Campus UIB, E07122 Palma de Mallorca, Spain.

[2]Innaxis Foundation & Research Institute, Jose Ortega y Gasset 20, E28006 Madrid, Spain

[*]Email: pfleurquin@ifisc.uib-csic.es

[*]Telephone: +34 971 25 98 82 / Fax: +34 971 17 32 48

# CHARACTERIZATION OF DELAY PROPAGATION IN THE US AIR TRANSPORTATION NETWORK

**ABSTRACT**

Complex networks provide a suitable framework to characterize air traffic. Previous works described the world air transport network as a graph where direct flights are edges and commercial airports are vertices. In this work, we focus instead on the properties of flight delays in the US air transportation network. We analyze flight performance data in 2010 and study the topological structure of the network as well as the aircraft rotation. The properties of flight delays, including the distribution of total delays, the dependence on the day of the week and the hour-by-hour evolution within each day, are characterized paying special attention to flights accumulating delays longer than 12 hours. We find that the distributions are robust to changes in takeoff or landing operations, different moments of the year or even different airports in the contiguous states. However, airports in remote areas (Hawaii, Alaska, Puerto Rico) can show peculiar distributions biased toward long delays. Additionally, we show that long delayed flights have an important dependence on the destination airport.

KEYWORDS: delay propagation, complex networks, Airport and Airline Performance, ATM.

## 1. INTRODUCTION

The generation, propagation and eventual amplification of flight delays involve a large number of interacting mechanisms. Such mechanisms can be classified as internal or external to the air traffic system. The basic internal mechanisms include aircraft rotations (the different flight legs that comprise an aircraft itinerary), airport operations, passengers' connections and crew rotation. In addition, external factors, such as weather perturbations or security threats, disturb the system performance and contribute to a high level of system-wide congestion. The intricacy of the interactions between all these elements calls for an analysis of flight delays under the scope of Complex Systems theory. Complexity is concerned with the emergence of collective behavior from the microscopic interaction of the system elements. Several tools have been developed to tackle complexity. Here we use Complex Networks theory and take a system-wide perspective to broaden the understanding of delay propagation. A network is a mathematical abstraction that represents systems of interacting entities as vertices (nodes) connected by edges (links) (see, for instance, Bocaletti et al. 2006, Newman 2010, or Barrat et al. 2012 for recent reviews). Given the natural networked structure of the air traffic system, we analyze the air transport network formed by nodes representing airports and edges direct flights between them. The nature of such network is highly dynamical since a different instance exists at every moment in time.

In this work we are interested in characterizing delays and how they may be transferred and amplified by subsequent operations, the so-called reactionary delays. Naturally reactionary delays spread across the network, so an understanding of the topological features of the air transportation network, the properties of aircraft rotations and the statistical features of flight delays is of great significance for subsequent modeling efforts (Fleurquin et al 2013).

The remainder of the paper is organized as follows. Section 2 provides a background review of the literature on complex networks, focusing on air transportation. Section 3 describes the used database. In Section 4 we present results on the characterization of the US air transportation network, flight trajectories and flight delays. Finally, Section 5 summarizes our findings and points to further research questions.

## 2. BACKGROUND

The use of network analysis to characterize complex systems has become widespread in the last two decades. The potential of graphs for describing social systems was pointed out almost a century ago (see Freeman 2004 for a review). However, the generalization of these concepts and tools had to wait much longer until the seminal works by Watts and Strogatz 1998 and by Barabási and Albert 1999. Ever since, complex networks have been applied in a growing range of disciplines such as technology (Huberman et al. 1999), biology (Jeong et al. 2001), or economy (Mantegna et al. 2007).

The application of network theory to air transportation has a much shorter history, for which the first results were published in 2004 and 2005. The world air transportation network is described as a graph formed with the passenger commercial airports as vertices and the direct flights between airports as edges (Barrat et al. 2004, Guimera et al. 2005), with a weight corresponding to the number of seats available in the connection. The main source of this database is IATA, while some other studies have presented data from the US Bureau of Transport Statistics (BTS) or from OAG. The initial work (Barrat et al. 2004) focused on the correlations between network topology and fluxes of passengers finding a non-linear relation between them: $w_{ij} = (k_i k_j)^\theta$, where $w_{ij}$ is the number of seats available in the connection between airports *i* and *j*, while $k_i$ is the number of connections with other airports of airport *i*, and $\theta$ is a parameter whose value was estimated to be approximately 1/2. A second study (Guimera et al. 2005) included a network description and analyzed the degree (number of connections per node) and node strength (sum over the weights of the connections of a node) distributions, degree-degree correlations, density of triangles, etc. The world air transportation network was analyzed later with graph clustering techniques (Sales-Pardo et al. 2007) to classify airports according to their connectivity patterns. The seasonal evolution of the connectivity patterns in the US airports networks have been also investigated in (Gautreau et al. 2009, Pan et al. 2011). The authors characterize along the year how the network connectivity varies, with more routes available in summer, as well as how the passenger fluxes modify. Recently, information on human mobility through the air transportation network has also been used to model and forecast the propagation pathways of infectious diseases transmitted by contact such as influenza (Balcan et al. 2009, 2009b).

Within the ATM community, even if reactionary delays have a great impact on air traffic performance (US Congress 2008, ICCSAI Fact Books 2011, Eurocontrol 2011), the research effort to understand delay propagation has been scarce so far, and mostly limited to a descriptive work (Beatty et al. 1999, Schaefer et al. 2001 & 2003, Ahmadbeygi et al. 2008). A good review of previous work on delay propagation can be found in (Belobaba et al. 2009, Jetzki 2009). Some research efforts have begun to apply network theory (Wuellner et al. 2010, Bonnefoy et al. 2007) in combination with stochastic modeling (Rosenberger 2002, Janic 2005) to the modeling of delay propagation (Bonnefoy et al. 2005, EPISODE 3, Pyrgiotis et al. 2013).

## 3. DATA & METHOD

Data was obtained from the Airline On-Time Performance Data available at the Bureau of Transportation Statistics webpage (www.bts.gov). This database provides information such as schedule and actual departure and arrival times, departure and arrival delays, origin and destination airports, taxi-in and taxi-out times, airline ID, tail number and flight date. Air carriers that exceed one percent of the total domestic scheduled-service passenger revenue, report on-time data and the causes of delay.

We restricted our analysis to domestic flights conducted in the year 2010. Despite these data are 2 years old, no major changes concerning on-time performance has occurred since then. For the year 2010, 18 air carriers filed on-time performance data that combined represents 6,450,129 flights from 305 airports. From this database 1.75% were cancelled and 0.2% diverted. All scheduled domestic flights for the year 2010 (not only those from On-Time Performance Data) totalize 8,687,800 (BTS 2011), therefore the data used represent 74% of all scheduled flights in 2010.

## 4. RESULTS

### 4.1 Characterizing the United States air transportation network

The resulting air transportation network is composed of 305 nodes denoting airports and 2,318 edges accounting for direct connections between them (Figure 1). Airports are sized according to the logarithm of their average delay per flight. Even though the network is not completely bidirectional, *i.e.*, there can be flights from A to B but not from B to A, most connections bear flights in the two directions. For example, we find that if we build daily networks with the information of the flights, 98% percent of the overall connections are bidirectional. Furthermore, the lowest percentage of bidirectional links measured in a daily network is 92%. Small airports are responsible for these minor anomalies. To simplify the analysis we symmetrized the network.

As in previous works, we define as degree of an airport its number of different connections (airports of origin or destination of flights connecting with it). We can then calculate a degree distribution taking into account the degrees or the number of flights of the airports across the network and integrate it to obtain a cumulative distribution $F_X(x)$, which for each value of $x$ is telling us which is the fraction of airports with degree (number of flights) less than or equal to $x$. In Figure 2, we show the complementary cumulative distribution of the degree and of the number of flights $(1 - F_X(x))$. Both distributions are wide and evince the heterogeneities present in the network. Some few airports are large hubs with a large number of connections and flights, while most of the airports have low traffic. These topological characteristics are well known for this network but still are relevant for the dynamics of delay propagation.

Table 1 shows the ranking of the top 10 airports based on the number of different destinations (degree) and displays also the number of flights. The largest hub in the network is Atlanta International Airport (ATL) with 159 direct connections and the average degree of the whole network is 15.2.

### 4.2 Flight trajectories

An important ingredient to characterize the propagation of reactionary delays is the rotation of the aircrafts. The database contains the tail number of the planes, which allows us to track their movements throughout the day. In Figure 3, we show the percentage of aircrafts taking a certain number of leaps per day. It can be seen that 80% of trajectories are composed of a number of leaps between 2 and 7. Very few planes do longer rotations due to the constraint of daily time periods and the duration of the flights.

Within a day, some of the aircraft trajectories for *closed walks*, that is, a sequence of airports starting and ending at the same airport, but most of the aircraft trajectories do not close at the end of the day. In Figure 4 we show the percentage of closed walks per day during 2010. We can conclude that these trajectories are a small percentage with respect to the total number of aircraft rotations. This finding does not mean that the trajectories will not close taking into account longer periods of time (weeks, months or years).

Regarding the previous result, another way of classifying the airports (besides connectivity) is according to the fraction of closed walks that starts in each airport. These airports are not necessarily the ones with highest degree (see Figure 5). Assuming that the airline hubs (airlines' centers of operations) are those airports with a larger percentage of closed rotations, we can conclude that the network hubs (nodes with highest degree) do not always coincide with the airlines hubs.

### 4.3 Flight delay characterization

We have described the topology of the network and the rotation of the flights. The next step is to focus on the real data regarding flight delays. We plot in Figure 6 the complementary cumulative distribution of departure and arrival delays for all flights of 2010, $1 - F_X(x)$. First, we notice that just like the degree and flight distribution, the delay distribution is broad with a slight hump at values of the delay around and larger than 700 min. Second, we find that there is no significant difference for both types of delays (arrival and departure delays), the day of the week or the season of the year (Figure 7). The cumulative distribution for different airports (Figure 8) shows a broad variety of behaviors. A remote airport from the mainland like Honolulu International Airport (HNL) and two continental hubs are displayed in the Figure: Dallas/Fort Worth International Airport (DFW) and Denver International Airport (DEN). We can see that DFW and DEN still show a slight hump in the distribution unlike HNL. On the other hand, Honolulu displays a broader distribution. This is probably due to the longer duration of the flights with destination or origin in HNL that allows for an easier absorption of short delays. The delays in the islands can be, therefore, much larger than those in the continent and as a consequence the distribution becomes more skewed.

In order to understand the nature of the hump in the delay distributions, we extract the flights with departure delay above 12 hours and compare them with all the flights of 2010. Plotting the departure delay as a function of the scheduled departure time we can distinguish how flights with delay greater than 12 hours are more abundant than the baseline at the beginning and at the end of the day (see Figure 9A). The opposite behavior can be observed for flights with departure delay below 12 hours, which show an almost flat delay distribution. Regarding this point, we plotted the delay distribution for flights with different scheduled departure times in Figure 9B. The hump becomes more evident in the distribution of flights departing between 00am to 5am and 1pm to 11:59pm (local times) indicating a relatively higher abundance of long delayed flights. Note that even so, the fraction of delay flights is small compared with the total.

Another feature of long delayed flights is their strong dependence on the destination airport. In Table 3, we compare the data for long delayed flights with two sets of randomly selected flights: one among all flights (delayed or not) and the other only with delayed flights. From the data 51 airports (16.00%) are the destination of 414 delayed flights. If the 414 flights are randomly chosen, the number of destination airport increases up to 120 (more than double the results from the real data) regardless of the way we choose the flights. This means that a bias exists towards a smaller set of destination airports. Note that the same phenomenon is not observed for the departure airports that are in the same range both in the data and in the randomly selected flights. Other variables as days, tail-number or air carriers remain the same. In Figure 10, we plotted the number of flights with long delays versus the ranking of destination airport with respect to the number of long delayed flights. The data correspond to the blue bars while the randomly selected set of flights are the red curve. In the data, the first 8 airports are destination of 75 % of the long delayed flights, while in the randomly selected set the first 8 airports totalize only 52 %.

The significance of the destination airports could be related to Ground Delay Program (GDP) from the Federal Aviation Administration (FAA). This program is implemented to control air traffic volume to airports where the estimated demand is expected to surpass the Airport Arrival Rate. When a GDP is issued flights destined to the affected airport are not permitted to depart until their Controlled Departure Time.

5. CONCLUSION

In summary, we have analyzed the characteristics of the US air transportation network with a focus on flight delays. The air transportation network is built by connecting pairs of airports if they have a direct flight. We studied the network topological properties such as the distribution of the number of flights or the number of connections per airport. These features show the broad heterogeneity of the air transport network in accordance with previous works. In addition to the topology, we consider also the properties of the aircraft rotation throughout the day and the characteristics of the delays. The aircraft rotation shows a complicated and highly heterogeneous profile. Some aircrafts itineraries are essentially round trips while others do not close in a simple periodic way. The heterogeneity of the rotation procedures can play an important role in the development and propagation of delays.

Regarding the delays, we show that the delay distributions show long decays both for arrival and departure delays, irrespective of the day of the week and season. Long tails are usually indicative of the complex nature of the mechanisms contributing to the propagation of delays. In this case, the system is not necessarily working under critical conditions but the combined action of several factors such as connecting passengers or crew, a predetermined schedule and the geographical distance of the airports can contribute to reach a similar system state at a global level. Whether the air transport network is a system at criticality is an open question that deserves further research. We study also the properties of the flights with a delay higher than 12 hours showing a relative concentration of long delayed flights early in the morning or late in the afternoon. The destination airport seems to be a key player for the surge of flights with long delay.

These results are relevant in order to better characterize flight delays from a statistical perspective. Subsequent efforts aimed at modeling delay spreading in the air transport networks, such as the recent works in (Fleurquin et al. 2013, 2013b), should have into account the statistical patterns described here both in the model development and validation.

**ACKNOWLEDGEMENTS**

PF is funded by the PhD program of the Complex World network of the WPE of SESAR. JJR is funded by the Ramón y Cajal program of the Spanish Ministry of Economy and Competitiveness (MINECO). Partial support was also received from MINECO (Spain) and FEDER (EU) through the project MODASS (FIS2011-24785) and from the EU Commission through the FP7 projects EUNOIA and LASAGNE.

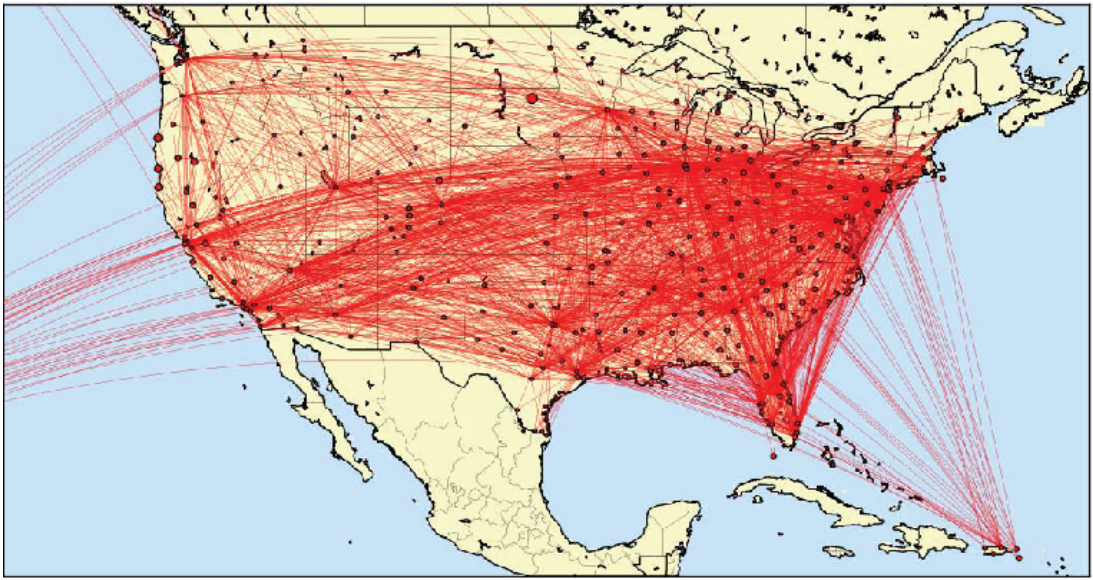

**FIGURE 1**

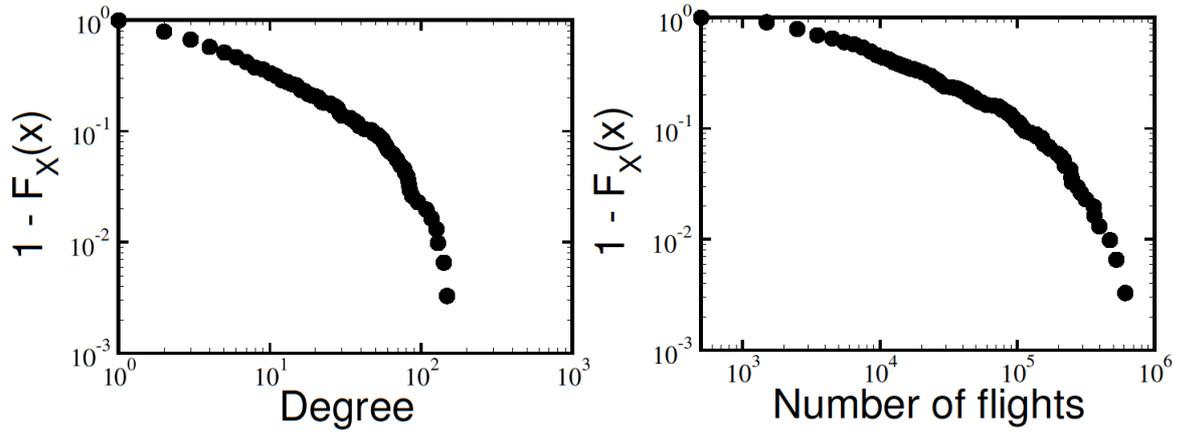

**FIGURE 2**

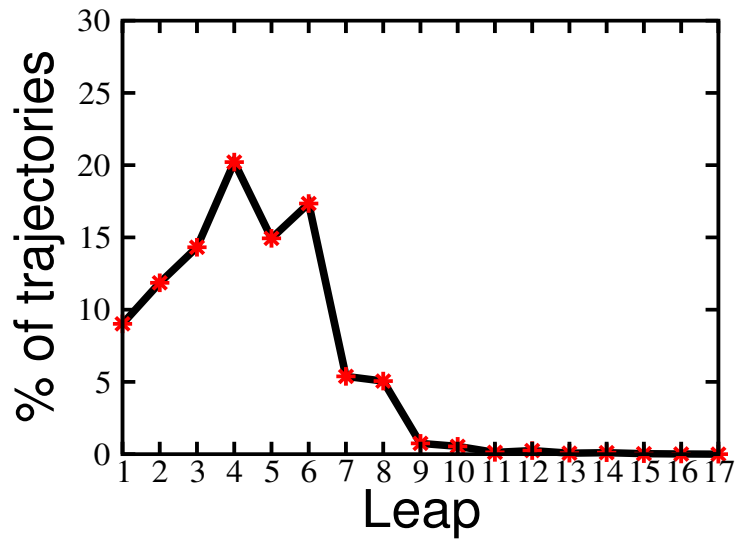

**FIGURE 3**
Pablo Fleurquin, José J. Ramasco, Víctor M. Eguíluz                              18

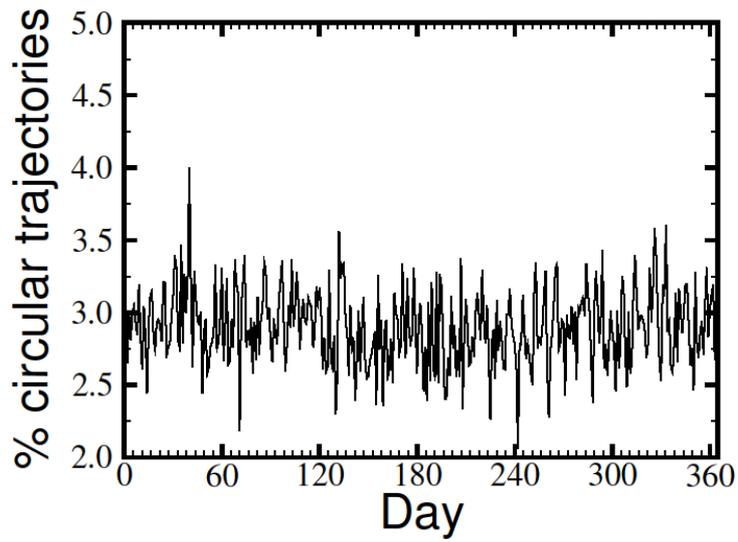

**FIGURE 4**

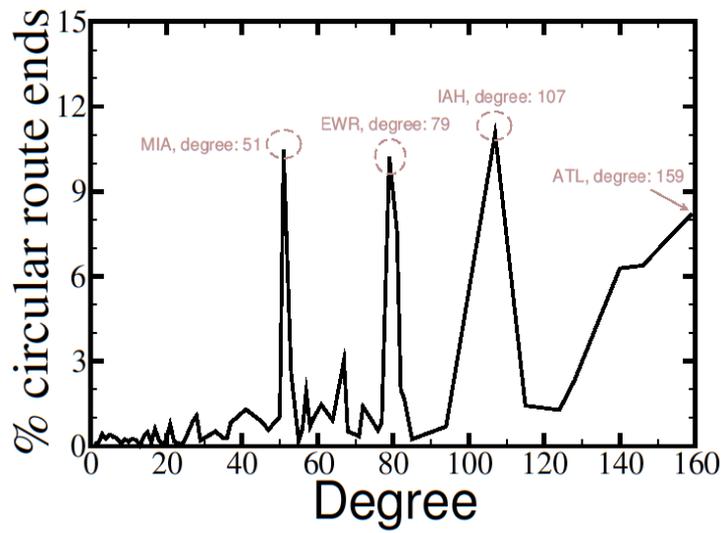

**FIGURE 5**

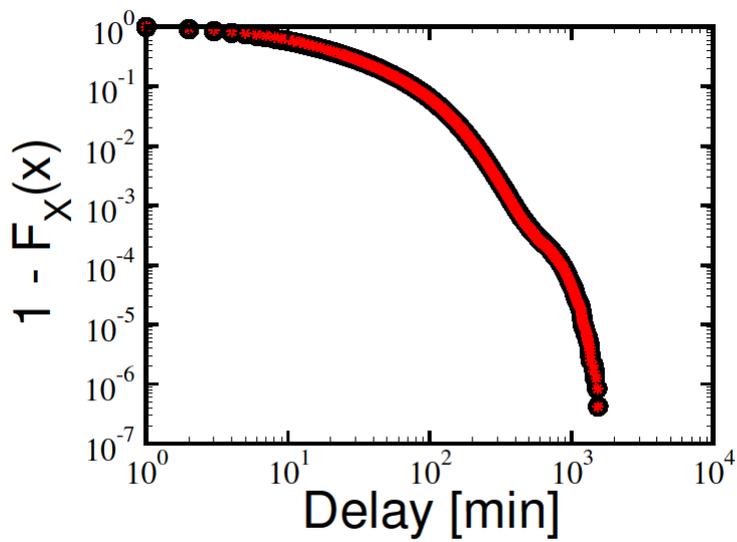

**FIGURE 6**

Pablo Fleurquin, José J. Ramasco, Víctor M. Eguíluz                                    19

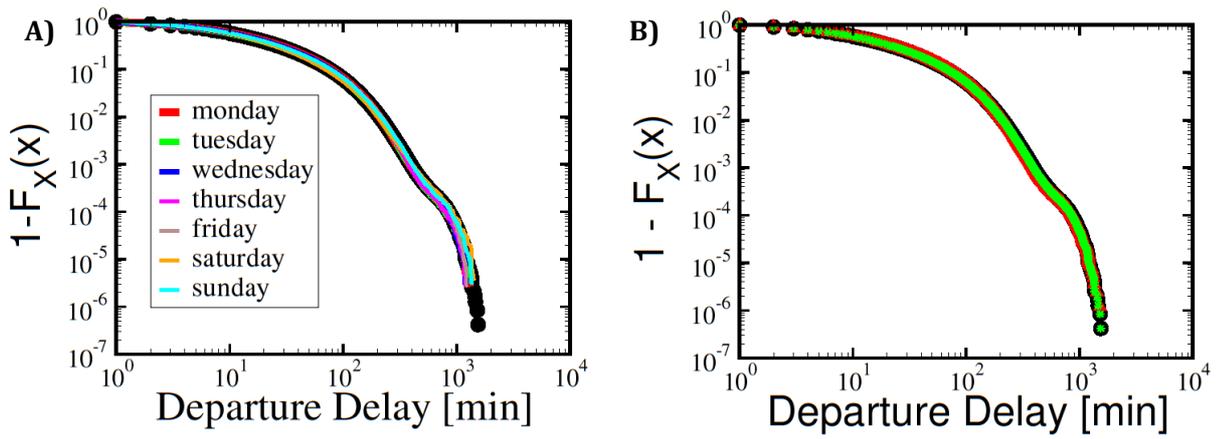

FIGURE 7

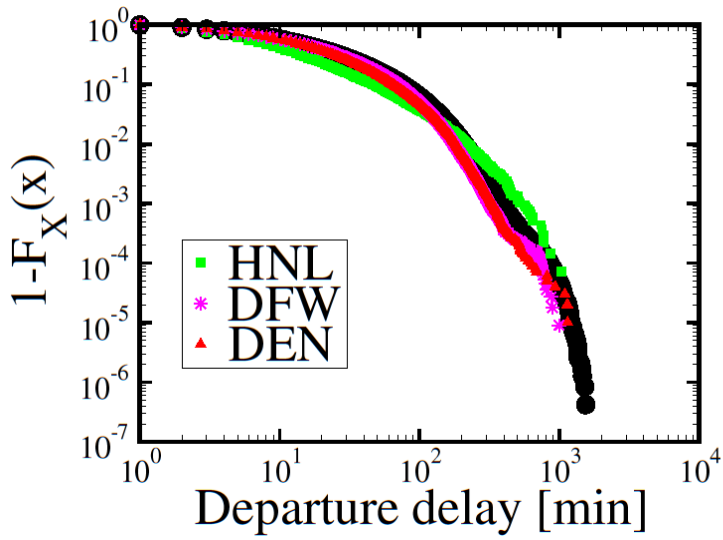

FIGURE 8

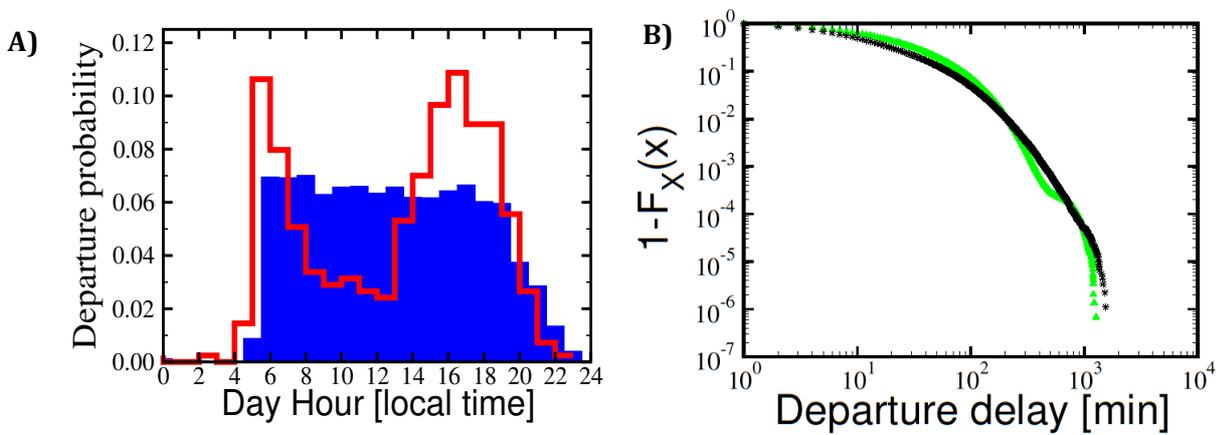

FIGURE 9

Pablo Fleurquin, José J. Ramasco, Víctor M. Eguíluz    20

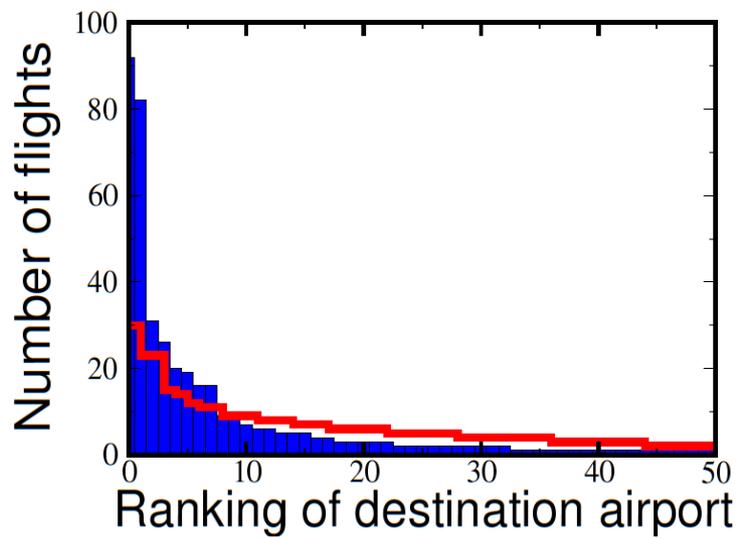

**FIGURE 10**

| Airport | # flights | Degree |
|---------|-----------|--------|
| ATL | 809,869 | 159 |
| ORD | 608,981 | 147 |
| DFW | 524,206 | 140 |
| DTW | 314,369 | 128 |
| DEN | 470,592 | 125 |
| MSP | 246,416 | 116 |
| IAH | 362,562 | 107 |
| SLC | 246,245 | 94 |
| MEM | 152,730 | 86 |
| MCO | 241,851 | 83 |

**TABLE 1**

| FLIGHTS WITH DEPARTURE DELAY >= 12 HOURS | | | | | | |
|---|---|---|---|---|---|---|
| | FLIGHTS | ORIGIN | DEST | DAYS | TAIL | AR_ID |
| WITH PROBLEM | 414 | 118 | 51 | 226 | 346 | 14 |
| TOTAL | 6341340 | 305 | 305 | 365 | 5081 | 18 |
| PERCENTAGE | 0.01% | 38.00% | 16.00% | 62.00% | 7.00% | 77.00% |
| **RANDOMLY CHOSEN 414 FLIGHTS** | | | | | | |
| | FLIGHTS | ORIGIN | DEST | DAYS | TAIL | AR_ID |
| WITH PROBLEM | 414 | 114 | 120 | 248 | 392 | 18 |
| TOTAL | 6341340 | 305 | 305 | 365 | 5081 | 18 |
| PERCENTAGE | 0.01% | 38.00% | 39.00% | 67.00% | 8.00% | 100.00% |
| **RANDOMLY CHOSEN 414 FLIGHTS DELAYED** | | | | | | |
| | FLIGHTS | ORIGIN | DEST | DAYS | TAIL | AR_ID |
| *RANDOMIZE* | 414 | 112 | 120 | 246 | 383 | 18 |
| TOTAL | 6341340 | 305 | 305 | 365 | 5081 | 18 |
| PERCENTAGE | 0.01% | 36.00% | 39.00% | 67.00% | 7.00% | 100.00% |

**TABLE 2**

**FIGURE 1**. US air transport network in 2010.

**FIGURE 2**. Complementary cumulative distribution of the degree, that is, the number of destinations per airport (on the left) and of the number of flights per airport (on the right) in the US air transport network.

**FIGURE 3**. Percentage of daily aircraft trajectories with given number of leaps.

**FIGURE 4.** Percentage of closed walks, that is, daily aircraft trajectories that start and end in the same airport.

**FIGURE 5.** Percentage of daily aircraft trajectories ending at an airport as a function of the airport degree. IATA codes are: MIA (Miami), EWR (Newark), IAH (Houston) and ATL (Atlanta).

**FIGURE 6.** Complementary cumulative distribution function of departure (black circles) and arrival (red stars) delay in 2010.

**FIGURE 7.** Complementary cumulative distribution function of departure delays in 2010. In A), the continuous colour lines represent data differentiated by weekday and black circles for all flights of 2010. In B), green stars correspond to flights operated in winter, red triangles represent flights operated in summer and black circles for all flights of 2010.

**FIGURE 8.** Complementary cumulative distribution of the departure delays in 2010 (black circles), and single airports HNL (Honolulu International Airport). DFW (Dallas Fort Worth) and DEN (Denver International Airport).

**FIGURE 9.** Fraction of departures as a function of the scheduled departure hour. A) Fraction of departures taking into account all flights of 2010 (blue bar), and fraction of departures per hour for flights with 12 hours departure delay or longer (red). B) Complementary cumulative distribution function of departure delays. Green triangles represents flights with scheduled departure from 00:00 am to 05:00 am or 01:00 pm to 11:59 pm. Black symbols represent flights with scheduled departure from 05:00 am to 00:59 pm.

**FIGURE 10.** Ranking of the number of flights delayed 12 hours or more for the 51 destination airport from the data (blue bars) and the randomly selected airports (red line). For the sake of clarity, from the 120 destination airports from the random case we only plot the first 51 airports.

**TABLE 1.** Major airports ranked according to their degree.

**TABLE 2.** Statistical analysis of flights with departure delay larger than 12 hours. For the year 2010 414 flights were delayed 12 hours or more. For comparison purposes 414 (delayed) flights were randomly selected, checking if this random selection modifies the origin or destination airport and also the number of days, aircrafts (tail number) or airline (airline id). Only the destination airport suffers a significant deviation from a random selection.